\documentstyle[preprint,epsf,aps]{revtex}
\begin{document}
\title{  Properties of a beam splitter entangler with Gaussian input states}
\author{Wang Xiang-bin\thanks{email: wang$@$qci.jst.go.jp} 
\\
        Imai Quantum Computation and Information project, ERATO, Japan Sci. and Tech. Corp.\\
Daini Hongo White Bldg. 201, 5-28-3, Hongo, Bunkyo, Tokyo 113-0033, Japan}

\maketitle 
\begin{abstract}
 An explicit formula is given
for the quantity of entanglement in the output state of a beam splitter, given the squeezed vacuum states input in each mode.
\end{abstract}
\section{Introduction}
As one of the few quantum devices that may act as the entangler, beam splitters have been extensively studied 
in its entangler related properties\cite{tan,sanders,paris,kim,kim1,agawal,campos,arvind,wang}.\\
In laboratory, coherent states and squeezed states are two practically existing robust states.
It is well known that no entanglement is produced if the input states are coherent states.
Therefore it is important to know the entanglement property when squeezed states are used as the
input. 
The output entanglement quantity is studied in ref\cite{kim} given the squeezed state input. 
In particular, an explicit formula expressing the output state in the form of two mode squeezed states
are given.
However, the result there is limited to a type of rather specific case. For example, the beam splitter
there is limited to the 50:50 beam splitter,  the input squeezed states can only have the real squeezing 
parameters and so on.
In this paper, we shall investigate this problem in a rather general background. We will give an explicit
formuly for the entanglement quantity of the output state.\\
It has been shown in ref\cite{arvind,wang} that in order to obtain an entangled output state,
 a necessary condition is that the input state should be non-classical. More generally, it was shown
in\cite{wang} that an arbitrary multi-mode classical state is still classical after an arbitrary
multi-mode rotation transformation. This means, for arbitrary linear optical system including
passive devices such as beam splitters, polarizing beam splitters, phase shifters, polarization
rotators and so on, the output multi-mode state must be classical( therefore separable) if the input is classical. However
However, this is only a necessary condition to obtain the entangled output state, it is
not a sufficient condition in general.  In certain case one may have interest to kno the exact amount of
entanglement in the output state of the beam splitter and how to maximize it through adjusting the parameters
in the passive linear optical system. Here we make an explicit calculation with the input of
two single mode squeezed states.\\
Consider a loseless beam splitter(see figure 1 in ref.\cite{wang}). We can distinguish the field  mode $a$ and mode
$b$ by the different propagating direction. Most generally,  the property of a beam splitter
operator $\hat B$ in Schrodinger picture can be summarized by the following equations(see e.g., ref\cite{campos0})
\begin{eqnarray}
\rho_{out}=\hat B \rho_{in} \hat B^{-1},
\end{eqnarray} 
\begin{eqnarray}
\hat B^\dagger=\hat B^{-1},
\end{eqnarray}
\begin{eqnarray}
\hat B \left(\begin{array}{c}\hat a\\\hat b\end{array}\right)\hat B^{-1}
=M_{B}\left(\begin{array}{c}\hat a\\ \hat b\end{array}\right)\label{m1},
\end{eqnarray}
\begin{eqnarray}
M_B
=\left(\begin{array}{cc}\cos\theta e^{i\phi_0}& \sin\theta e^{i\phi_1}\\ -\sin\theta e^{-i\phi_1}
& \cos\theta e^{-i\phi_0} \end{array}\right)
\label{m2}\end{eqnarray} 
\begin{eqnarray}
\hat B |00\rangle=|00\rangle\label{va}.
\end{eqnarray}
Here $\rho_{in}$ and $\rho_{out}$ are the density operator for the input 
and output states respectively. Both of them are two mode states including mode $a$
and mode $b$. The elements in the matrix $M_B$ 
are determined by the beam splitter itself, 
$\hat a,\hat b$ are the annihilation operators for  mode $a$ and mode $b$
respective, $|00\rangle$ is the vacuum
state for both mode.  Equation (\ref{va}) is due to the simple fact of no input no output.\\
\section{Inseparability quantity with squeezed states input } Suppose the input states are the squeezed vacuum states in each mode, i.e. 
\begin{eqnarray}
\rho_{in}=\hat S_a(\zeta_a)\hat S_b(\zeta_b)|00\rangle\langle 00| \hat S^\dagger(\zeta_a)\hat S^\dagger(\zeta_b),
\end{eqnarray} 
where
\begin{eqnarray} \begin{array}{l}
\hat S_a(\zeta_{a})=\exp\left(\frac{1}{2}\zeta_a^{*}\hat a^2-\frac{1}{2}\zeta_a \hat a^{\dagger 2}\right);\\
\hat S_b(\zeta_{b})=\exp\left(\frac{1}{2}\zeta_b^{*}\hat b^2-\frac{1}{2}\zeta_b \hat b^{\dagger 2}\right).
\end{array}
\end{eqnarray}
They have the following propertis
\begin{eqnarray}
\hat S_a^\dagger(\zeta_a)\left(\begin{array}{c}a^\dagger\\a\end{array}\right)
\hat S_a(\zeta_a)=\left(\begin{array}{cc}
\cosh r_a & -e^{-i\chi_a}\sinh r_a\\-e^{i\chi_a}\sinh r_a & \cosh r_a
\end{array}\right)\left(\begin{array}{c}b^\dagger\\b\end{array}\right)
;\label{s1}
\end{eqnarray}
\begin{eqnarray}
\hat S_b^\dagger(\zeta_b)\left(\begin{array}{c}b^\dagger\\b\end{array}\right)
\hat S_b(\zeta_b)=\left(\begin{array}{cc}
\cosh r_b & -e^{-i\chi_b}\sinh r_b\\-e^{i\chi_b}\sinh r_b & \cosh r_b\end{array}\right)  
\left(\begin{array}{c}b^\dagger\\b\end{array}\right)
,\label{s2}
\end{eqnarray}
where
$r_{a,b}=|\zeta_{a,b}|$ and  $\chi_{a,b}=\tanh^{-1}\zeta_{a,b}/r_{a,b}$.

For simplicity, we use the characteristic function for the input state $\rho_{in}$ and the output state $\rho_{out}$.
$$
C_{in} (\xi_a,\xi_b ) 
={\rm tr} \left[ \exp\left (\xi_a \hat a-\xi_a^* \hat a^\dagger
+\xi_b \hat b-\xi_b^*\hat b^\dagger \right) \rho_{in} \right] $$
\begin{eqnarray}
={\rm tr} \left\{ \exp \left[i\sqrt 2(\xi_a^I \hat x_a+\xi_a^R \hat p_a
+\xi^I_b \hat x_b+\xi_b^R\hat p_b \right)\rho_{in} \right\},
\end{eqnarray}
where the parameters $\xi_{a,b}=\xi_{a,b}^R+i\xi_{a,b}^I$, 
$ (\hat x_{a}, \hat p_{a})= N( \hat a^\dagger, \hat a)^T  $,  
$ (\hat x_{b}, \hat p_{b})= N( \hat b^\dagger, \hat b)^T $  and 
$N=\frac{1}{\sqrt 2}\left(\begin{array}{cc}1 & 1\\-i&i\end{array}\right)$. 
For convenience, we denote $\hat D(\xi_a,\xi_b)=\exp\left (\xi_a \hat a-\xi_a^* \hat a^\dagger
+\xi_b \hat b-\xi_b^*\hat b^\dagger \right) $
In the case of squeezed states input, the characteristic function  for the output state is
$$
C_{out}(\xi_a,\xi_b)={\rm tr}\left[ \hat D(\xi_a,\xi_b)\hat B \hat S_a(\zeta_a)\hat S_b(\zeta_b)|00\rangle\langle 00| 
\hat S_a^\dagger(\zeta_a)\hat S_b^\dagger(\zeta_b)\hat B^{\dagger} \right]$$\begin{eqnarray}
={\rm tr}\left[ \hat S_a^\dagger(\zeta_a)\hat S_b^\dagger(\zeta_b) B^\dagger \hat D(\xi_a,\xi_b)\hat B 
\hat S_a(\zeta_a)\hat S_b(\zeta_b)|00\rangle\langle 00| 
\right].\label{trans}\end{eqnarray}
Suppose the output state of mode a being $\rho_{oa}$. The quantity of entanglement for the output state
between mode $a$ and mode $b$ is
\begin{eqnarray}
E(\rho_{oa})={\rm tr}(\rho_{oa}\ln\rho_{oa})\label{ed}.
\end{eqnarray}
Using eq(\ref{trans}), we can calculate the characteristic function for the output state in mode $a$ explicitly:
$$
C_{oa}(\xi_a)=C_{out}(\xi_a,\xi_b=0)
=\exp\left[-\frac{1}{2}\cos^2\theta \left|\xi_a^* e^{i\phi_0}\cosh r_a
+\xi_a  e^{-i\phi_0+i\chi_a} \sinh r_a\right|^2\right]
$$
\begin{eqnarray}
\cdot\exp\left[-\frac{1}{2}\sin^2\theta\left|\xi_a e^{-i\phi_1}\cosh r_b
+\xi_a^* e^{i\phi_1-i\chi_b} \sinh r_b\right|^2
\right].\label{cha1}
\end{eqnarray}
In obtaining the above equation, we have used equation(\ref{m1},\ref{m2}), equation(\ref{s1},\ref{s2}) to reduce the part
$\hat S_a^\dagger(\zeta_a)\hat S_b^\dagger(\zeta_b) B^\dagger \hat D(\xi_a,\xi_b)\hat B 
\hat S_a(\zeta_a)\hat S_b(\zeta_b)$.
The right hand side of equation(\ref{cha1}) can be written in the form in $\xi_R$ and $\xi_I$ where $\xi_R +i\xi_I=\xi_a$, i.e.
\begin{eqnarray}
C_{oa}=\exp\left[-\frac{1}{2}(\xi_R,\xi_I)M_{oa}(\xi_R,\xi_I)^T\right].
\end{eqnarray}
Here $M_{oa}$ is the $2\times 2$ covariance matrix as 
$M_{oa}= \left(\begin{array}{cc} m_{11} & m_{12}\\
m_{21} & m_{22}\end{array}\right).$
After calculation we obtain the matrix elements  
\begin{eqnarray}
m_{11}=\Sigma_a\cos^2\theta+\Sigma_b\sin^2\theta+2x_a\cos^2\theta\cos\Delta_a+2x_b\sin^2\theta\cos\Delta_b; 
\label{mm1}\end{eqnarray}
\begin{eqnarray}
m_{12}= m_{21} = 
2x_a\cos^2\theta\sin\Delta_a+2x_b\sin^2\theta\sin\Delta_b;
\label{mm2}\end{eqnarray}
and
\begin{eqnarray}
m_{22}=\Sigma_a\cos^2\theta+\Sigma_b\sin^2\theta-2x_a\cos^2\theta\cos\Delta_a-2x_b\sin^2\theta\cos\Delta_b, 
\label{mm3}\end{eqnarray}
where $\Sigma_a=\cosh^2 r_a +\sinh^2 r_a$, $\Sigma_b=\cosh^2 r_b +\sinh^2 r_b$, 
$x_a=\sinh r_a\cosh r_a$, $x_b=\sinh r_b\cosh r_b$, $\Delta_a=2\phi_0-\chi_a$ and 
$\Delta_b=2\phi_1-\chi_b$. 
We can choose an appropriate unitary transformation to $\rho_{oa}$ to obtain another
density operator $\rho_{oa}'$ whose characteristic function is
\begin{eqnarray}
C_{oa}'(\xi_a)=\exp\left[-\frac{1}{2}(\xi_R,\xi_I)\left(
\begin{array}{cc}\delta & 0\\0 & \delta\end{array}\right)(\xi_R,\xi_I)^T\right]\label{d}
\end{eqnarray}  
and 
\begin{eqnarray}
\delta=\sqrt{m_{11}m_{22}-m_{12}^2}.\label{delta}
\end{eqnarray}
We know the Wigner characteristic function for a thermal state 
$ (1-e^{-\beta})e^{-\beta a^\dagger a} $ is \cite{op}
\begin{eqnarray}
C_{th}(\xi)=\exp\left[-\frac{1}{2} (\xi_R,\xi_I) \left(\begin{array}{cc}
\frac{1+e^{-\beta}}{1-e^{-\beta}} &0\\0& \frac{1+e^{-\beta}}{1-e^{-\beta}} 
\end{array}\right) (\xi_R,\xi_I) \right],
\end{eqnarray}
This is to say, the state defined by the characteristic function in equation(\ref{d}) is a thermal state 
in the form 
\begin{eqnarray}
\rho_{oa}'=(1-e^{-\beta})e^{-\beta a^\dagger a} 
\label{ths}\end{eqnarray}
with the parameter $\beta$ satisfying
\begin{eqnarray}
e^{-\beta}=\frac{\delta-1}{\delta +1}.
\end{eqnarray}
Since the trace value does not change under any unitary transformation, the entanglement quantity defined in 
equation(\ref{ed}) is
\begin{eqnarray}
E(\rho_{oa})={\rm tr} \rho_{ao}'\ln\rho_{ao}' 
\end{eqnarray}For the thermal state
defined by equation(\ref{ths}), calculation for the quantity ${\rm tr} \rho_{ao}'\ln\rho_{ao}'$ is straightforward.
Thus we have the following result for the quantity of entanglement for the output state given the squeezed state input in each mode:
\begin{eqnarray}
E(\rho_{out})=\ln(1-e^{-\beta})+\frac{\beta e^{-\beta}}{1-e^{-\beta}}=\ln\frac{2}{\delta+1}-
\frac{\delta-1}{2}\ln\frac{\delta-1}{\delta+1},
\end{eqnarray} 
with $\delta$ being defined by equation(\ref{delta}) and equation(\ref{mm1},\ref{mm2},\ref{mm3}). The above equation together
with the previous equations for the definition of $\delta$ gives a direct calculation formula for the entanglement quantity
given the indepedent squeezed state as the input to each mode. This is to say, the maximum value of $\det M_{oa}$ gives the
largest entanglement. In order to maximize the entanglement, we should maxmize the value of
$\delta$. After calculation we can see that
\begin{eqnarray}
\delta^2=
(\Sigma_a+\Sigma_b+ \sinh 2r_a +  \sinh 2r_b)(\sin^4\theta+\cos^4\theta)+\frac{1}{2}\Sigma_a\Sigma_b \sin^2 2\theta -2x_ax_b\sin^2 2\theta\cos(\Delta_b-\Delta_a)
\end{eqnarray}
Obviously the following condition is required to maximize the value of $\delta^2$ for the maximum entanglement
\begin{eqnarray}\label{cond}
\Delta_b-\Delta_a=2(\phi_1-\phi_0)-(\chi_b-\chi_a)=(2k+1)\pi,
\end{eqnarray}
where $k$ is an arbitrary integer. And we know that, the values of
both $\chi_b-\chi_a$ and $\phi_1+\phi_0$ are practically detectable and controllable
in a beam splitter experiment. 
This constraint is independent of $\theta$ or $r_a,r_b$. In particular, taking the special case 
$\phi_0$ and $|\cos\theta|=1/2$ it is just the result
 given by Kim et al\cite{note}. However, our result is more general than that in ref\cite{kim}.
Ref.\cite{kim} has only given the maximum point in the case of 50:50 beam splitter with $\phi_0=0$. 
No explicit formula for the quantity of entanglement is given there\cite{kim}. Our result is more general in that it can not only be used 
for the exact amount of entanglement but also to find the maximum point of 
entanglement
for the output state of 
a beam splitter with arbitrary transmission rate and with arbitrary 
phase values of $\phi_0,\phi_1,\chi_a$ and $\chi_b$.
\section{Concluding remark} In summary, we have studied the entanglement quantity for the output state of a beam splitter given 
 squeezed vacuum state as the input state in each mode. 
Different from the previous result\cite{kim}, our result is not limited to
the 50:50 result. We don't know how to obtain the more general result
given the general Gaussian state, since so far there is no good entanglement
for the impure Gaussian state. It has been shown in ref\cite{kim} that nonclassical
separable input state can be changed to an entangled state in the output. The inverse  
of such a process makes examples that even though the input state is nonclassical, the 
output could be still separable. Some specific examples are given in \cite{campos}. The
necessary and sufficient condition for an inseparable output state is not given so far. 
It is possible to obtain the necessary and sufficient
condition for inseparability of the output state given the Gaussian input state. We will
give this condition explicitly in this paper.However, one may still easily find the
criterion on whether the output state is inseparable through  
inseparability criterion\cite{werner}:
\begin{eqnarray}
M_{out}+ i\widetilde \sigma \geq 0,
\end{eqnarray} 
where M is the correlation matrix of the output state, 
the $4\times 4$ matrix $\widetilde \sigma =J_A^T\oplus J_B$,  $J_A=J_B=\left(\begin{array}{cc}0&-1\\1&0\end{array}\right)$. 
 A detailed calculation on this is given in ref\cite{qw}.  
\\{\bf Acknowledgement:} I thank Prof Imai for support. I thank Dr Hwang WY, Dr Winter A, Dr Yura H, Dr Matsumoto K,and Dr Tomita A for useful
discussions.


\begin{thebibliography}{99}
\bibitem{note} To see this point, we should first observe eq.(13) in ref\cite{kim}. In our paper's notation, eq.(13) of ref\cite{kim} reads $\hat B(\theta,\phi_1)\hat S_a(\zeta_a)\hat S_b(\zeta_b)|00\rangle = \hat R(-\chi_a/2)\hat R (\chi_b/2)\hat B(\theta, \phi_1-\chi_b/2+\chi_a/2)\hat S_a(r_a)\hat S_b(r_b)|00\rangle$ with the special value of
$\phi_0=0$. The condition ``$\phi=\pi/2$'' in ref\cite{kim} reads
$\phi_1-\chi_b/2+\chi_a/2=\pi/2$ in our notation. This is in agreement with
eq(\ref{cond}) in this paper.
\end{thebibliography}
\end{document}